\documentclass[prl,aps,twocolumn,groupedaddress,floats,showpacs,final,superscriptaddress]{revtex4}
\usepackage{graphicx}
\usepackage{bm}
\usepackage{color}
\usepackage{epsfig}
\usepackage{psfrag}
\usepackage{amsmath,amsfonts,amssymb,bm,ulem}

\begin{document}

%%%%%%%%%%%%%%%%%%%%%%%%%%%%%%%%%%%%%% AUTHORS %%%%%%%%%%%%%%%%%%%%%%%%%%%%%%%%%%
\author{Igor S. Tupitsyn}
\affiliation{Department of Physics, University of Massachusetts, Amherst, MA 01003, USA}
\affiliation{National Research Center ``Kurchatov Institute'', 123182 Moscow, Russia}
\author{Nikolay V. Prokof'ev}
\affiliation{Department of Physics, University of Massachusetts, Amherst, MA 01003, USA}
\affiliation{National Research Center ``Kurchatov Institute'', 123182 Moscow, Russia}
\affiliation{Department of Theoretical Physics, The Royal Institute of Technology,
Stockholm SE-10691 Sweden}

%%%%%%%%%%%%%%%%%%%%%%%%%%%%%%%%%%%%%%% TITLE %%%%%%%%%%%%%%%%%%%%%%%%%%%%%%%%%%%%

\title{Stability of Dirac Liquids with Strong Coulomb Interaction}
\date{\today}

%%%%%%%%%%%%%%%%%%%%%%%%%%%%%%%%%%%%%% ABSTRACT %%%%%%%%%%%%%%%%%%%%%%%%%%%%%%%%%%

\begin{abstract}
We develop and apply the Diagrammatic Monte Carlo technique to address the problem
of stability of the Dirac liquid state (in a graphene type system) against strong
long-range part of the Coulomb interaction. So far, all attempts to deal with this problem
in the field-theoretical framework were limited either to perturbative or RPA treatments,
with diametrically opposite conclusions. Our calculations aim at the approximations-free
solution with controlled accuracy by computing vertex corrections from higher-order
skeleton diagrams and establishing the renormalization group flow of the effective
Coulomb coupling constant. We unambiguously show that with increasing the system
size $L$ (up to $ln(L) \sim 40$), the coupling constant always flows towards zero;
i.e. the two dimensional Dirac liquid is an asymptotically free $T=0$ state with
divergent Fermi velocity

\end{abstract}


\maketitle
%%%%%%%%%%%%%%%%%%%%%%%%%%%%%%%%%%%%%% ARTICLE BODY %%%%%%%%%%%%%%%%%%%%%%%%%%%%%%%%%%%%

The linear in momentum low-energy part of electronic spectrum with vanishing
density of states at the Fermi points in the undoped graphene results in a picture
of massless two-dimensional (2D) Dirac fermions in a semimetallic state. While
conventional 3D metals efficiently screen long-range Coulomb interactions and
fall under the standard Fermi liquid description, 2D Dirac fermions leave
Coulomb interactions unscreened. This, in turn, leads to a divergent renormalization
of quasiparticle properties \cite{CastrNeto2012,DasSarma2007}, and the system is
commonly referred to as the Dirac liquid (DL).

\begin{figure}[tbh]
%\centering
\vspace{-1.2cm}
\includegraphics[width=0.75\columnwidth, angle=-90]{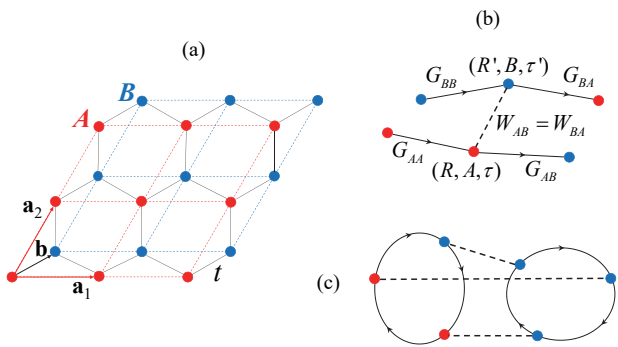}
\vspace{-1.1cm}
\caption{ (color online). (a) Honeycomb lattice of graphene with the nearest-neighbor
hopping amplitude $t$ ($|{\bf a}_1|=|{\bf a}_2|=a$). (b) The basic building block for the $G^2W$
skeleton expansion is based on three-point vertexes characterized by the unit cell index
${\mathbf R}$, imaginary time $\tau$, and sublattice index $\xi = \{ {\bf A},{\bf B}\}$. Green's
functions $G$ and screened interactions $W$ are connecting three-point vertexes pairwise. (c)
Typical skeleton free-energy diagram. }
\label{Fig1}
\end{figure}

The effect of long-range Coulomb interactions (unrelated to the transition to the
AFM insulator state with spontaneously broken chiral sublattice symmetry \cite{Semenoff1984}
caused by strong on-site repulsion) on properties of DL has been addressed
theoretically in a number of works using both analytic and numeric
approaches (see \cite{CastrNeto2012} and references therein). Within the lowest-order
perturbation theory (Fock diagram) \cite{Gonzales1994} it was found that the
effective coupling constant
$\alpha = e^2/\epsilon_0 v_F$ (where $e$ is the electron charge,
$\epsilon_0$ is the background dielectric constant, and $v_F$ is the Fermi velocity)
renormalizes to zero logarithmically as the system size $L$ is increased
\begin{equation}
d\alpha (l)/dl = -\alpha^2(l)/4 \;,  \;\;\;\; l=\ln(L/a) \, .
\label{RG1}
\end{equation}
Here $a$ is the lattice constant (see Fig.\ref{Fig1}(a)). In the absence of
charge renormalization \cite{Gonzales1994,Ye} this is equivalent to divergent renormalization
group (RG) flow for the Fermi velocity $dv_F(l)/dl = \alpha v_F/4 = const$;
i.e., the theory is asymptotically free if the bare coupling constant
$\alpha_0 = e^2/\epsilon_0 v_F^{(0)}$ is small ($v_F^{(0)}=\sqrt{3}\,at/2$
is the Fermi velocity in the non-interacting system).

However, in suspended graphene the bare Coulomb coupling constant is not small,
$\alpha_0 = 2.2$. When the RG equation for $v_F$ is computed up
to the next-to-leading order
in $\alpha$, it is found that the flow features an unstable infrared fixed
point \cite{Vafek2008}. At this level of approximation,
\begin{equation}
dv_F/dl = [1-c\alpha]\, \alpha v_F/4  \;,  \;\;\;\; c\approx 1.2 \,.
\label{RG2}
\end{equation}
and the flow is towards strong coupling if $\alpha_0 > \alpha_c \approx 0.8$.
On the one hand, this result hints at the possibility that DL may be unstable
against strong Coulomb interactions. On the other hand, the value of $\alpha_c$
is not small and the second-order perturbative result cannot be trusted.
Indeed, when
the second-order calculation is upgraded to include an infinite sum of
bubble diagrams (the so-called RPA approximation) the unstable infrared point is
removed \cite{Gonzales1999,Son2007,DasSarma2014}; the same conclusion was reached
within the functional renormalization group approach \cite{Kopietz2016}. However,
given that for $\alpha \ge 1$ all quantities are strongly renormalized, unaccounted
(higher-order) diagrammatic contributions may alter the final result, leaving the question of
stability of DL for strong Coulomb interactions an unsolved theoretical problem.

It should be mentioned that suspended undoped graphene is a semimetal, and
significant enhancement of the Fermi velocity observed in measurements of the cyclotron
mass \cite{Elias2011} and ARPES spectra \cite{Siegel2011} indicates that we are dealing
with stable DL. Early Hybrid Monte Carlo simulations of the effective two-band model of
graphene \cite{GraphNum1,GraphNum2,GraphNum3} with strong on-site repulsion and
$ \alpha_0 v_F^{(0)} /r$ Coulomb term for the rest of the lattice predicted an insulating
state for $\alpha_0 > 1$, in line with (\ref{RG2}) and in contradiction with experiments.
However, when a more realistic parametrization of inter-particle interactions at short
distances was introduced
\cite{Ulybyshev2013,Smith2014}, suspended graphene was found to remain semimetallic.
These results show that one has to be extremely careful in separating physics of
strong short-range correlations from the RG flow due to long-range forces.
Unfortunately, the largest system sizes simulated in Ref.~\cite{Ulybyshev2013}
($L/2a \le 12$) were too small for constructing the RG flow of the effective
Coulomb coupling.

In this Letter we develop the bold-line Diagrammatic Monte Carlo (BDMC)
technique for graphene type systems that allows us to deal with Coulomb interactions
in a fully self-consistent, approximations free, manner and obtain final results with
controlled accuracy by accounting for vertex corrections from higher-order skeleton diagrams.
To demonstrate that BDMC leads to an accurate solution, we first benchmark
the technique against a much harder (as far as the diagrammatic series are concerned)
problem of the semimetal--insulator transition in suspended graphene by comparing our results
with Refs.~\cite{Ulybyshev2013,Smith2014}. The main topic of this study, however, is
stability of DL against the strong long-range part of the Coulomb interaction,
and we were able to establish the RG flow of the effective coupling constant
over twelve(!) orders of magnitude in length scales. We find that the system
always flows to the asymptotically free DL state (to suppress short-range correlations
the interatomic potential is made constant at distances $r\le 2a$).
The proper solution of the problem does require that higher-order vertex corrections
are accounted for in a fully self-consistent way because
they significantly renormalize the flow at strong coupling.

%%%%%%%%%%%%%%%%%%%%%%%%%%%%%%%%%%%%%%%%%%%%%%%%%%%%%%%%%%%%%%%%%%%%%%%%%%%%%%
{\it System.} Carbon atoms in graphene are arranged in a honeycomb lattice that
can be seen as two identical triangular sublattices, ${\bf A}$ and {\bf B},
with unit vectors $\{{\bf a}_1,{\bf a}_2 \}$,
shifted relative to each other by ${\bf b}=({\bf a}_1+{\bf a}_2)/3$, see
Fig.\ref{Fig1}(a). In what follows we use $a$ as a unit of length. The Hamiltonian
$H =H_0+H_{\rm int}$ is defined by
\begin{eqnarray}
H_0 & =&
- t \sum_{<{\mathbf i} {\mathbf j}> \sigma}
(a^{\dag}_{{\mathbf i} \sigma} \;b^{\,}_{{\mathbf j} \sigma} + h.c.)
% &-& t'\sum_{<<{\mathbf i}{\mathbf j}>> \sigma}
% (a^{\dag}_{{\mathbf i} \sigma} a^{}_{{\mathbf j} \sigma}
% + b^{\dag}_{{\mathbf i},\sigma} b^{}_{{\mathbf j},\sigma} + h.c.)
- \mu \sum_{{\mathbf i} \sigma}  n_{{\mathbf i} \sigma}  \, \label{H0} \\
H_{\rm int} &=& \frac{1}{2} \sum_{{\mathbf i} {\mathbf j} \sigma \sigma'}
 V_{\sigma \sigma'}(\vert {\bf r}_{{\mathbf i}}-{\bf r}_{{\mathbf j}} \vert )
\; n_{{\mathbf i} \sigma} n_{{\mathbf j} \sigma'} \, ,
\label{Hint}
\end{eqnarray}
where $H_0$ is based on the standard tight-binding approximation characterized
by the nearest-neighbor (between sublattices) hopping amplitude $t$ and chemical potential
$\mu$. [BDMC technique can deal with arbitrary lattice dispersion relation.]
The second term describes electron-electron interactions with the Coulomb-law
form at large distances (we employ standard second-quantization notations
for creation, annihilation, and density operators in the site representation).
The on-site coupling, $V_{\sigma \sigma'}(0) =U\delta_{\sigma, -\sigma'}$,
is the only interaction term that depends on the spin index $\sigma=\pm $;
all other couplings are spin-independent,
$V_{\sigma \sigma'}(r > 0) = V_{c}(r)$ with $V_{c}(r>>a) \to e^2/\epsilon_0\,r $.

%%%%%%%%%%%%%%%%%%%%%%%%%%%%%%%%%%%%%%%%%%%%%%%%%%%%%%%%%%%%%%%%%%%%%%%%%%%%%%%%%%%

{\it Formalism}. Our calculations are based on the standard Feynman diagrammatic technique
re-formulated in terms of the self-consistent $G^2W$ skeleton expansion \cite{Heidin}.
In the real-space imaginary-time representation the free-energy skeleton diagrams
are composed of three-point vertexes located at space-time positions $({\bf r}, \tau)$
and connected pairwise by fully-dressed Green's functions, $G$, and screened  effective
interactions, $W$, see Fig.~\ref{Fig1}(b), in such a way that the resulting graph is
(i) connected, (ii) cannot be made disconnected by cutting two lines of the same kind,
see Fig.~\ref{Fig1}(c). The only exception is the Hartree diagram, which can be absorbed
into the chemical potential (in general, spin and sublattice dependent).

Within the BDMC framework (see Refs.~\cite{KULAGPRL2013,EPI2016} for more details),
the configuration space of skeleton diagrams for free-energy is sampled stochastically;
by removing one of the graph lines, either $G$ or $W$, one obtains a diagram either
for the proper electron self-energy $\Sigma$ or irreducible polarization function $\Pi$.
The self-consistency loop is closed by Dyson equations that take an algebraic form in the
momentum-frequency space:
\begin{equation}
G^{-1} = G_0^{-1}-\Sigma \;, \qquad \qquad  W^{-1} = V^{-1}-\Pi\, .
\label{Dyson}
\end{equation}
For brevity, we do not explicitly mention the tensor structure of interactions, propagators,
and irreducible objects in the sublattice and spin space. On a honeycomb lattice all
quantities are $2 \times 2$ matrices in the basis $| \phi_{\xi}({\bf R}) \rangle$, where
$\xi = \{ {\bf A},{\bf B}\}$ and ${\bf R}$ is the unit cell index. In this basis,
the bare Green's function is given by
$(G_0)_{\xi \xi'}({\bf R}-{\bf R}') = \sum_{\gamma}
\langle \phi_{\xi} ({\bf R}) | \Phi_{\gamma} \rangle
(G_0)_{\gamma} \langle \Phi_{\gamma} | \phi_{\xi'} ({\bf R}') \rangle$
with $(G_0)_{\gamma} = (i\omega_n + \mu -  \epsilon_{s}({\mathbf k}))^{-1}$, where
$| \Phi_{\gamma} \rangle$ is the $\gamma =(s, {\mathbf k})$-th eigenstate
of the tight-binding Hamiltonian with eigenenergy
$\epsilon_{s=1,2}({\mathbf k})$ and $\omega_n =
2\pi T (n+1/2)$ with integer $n$ is the fermionic Matsubara frequency.

Our implementation of the BDMC technique, generalized to several atoms in the unit cell,
closely follows that of Ref.~\cite{KULAGPRL2013}. Both $\Sigma$ and $\Pi$ are computed
as sums of all skeleton graphs up to order $N$ (there are $2N$ vertexes in the $N$-th
order graph); we denote these sums as $\Sigma_N$ and $\Pi_N$. The lowest-order
contributions $\Sigma_1$ and $\Pi_1$ are nothing but products of $G$ and $W$
functions, and at $N=1$ the scheme is identical to the GW-approximation.
Monte Carlo statistics has to be collected only from higher-order diagrams.
The skeleton formulation is complete and the diagrammatic sequence
for long-range Coulomb interactions on a lattice is expected to converge
with increasing the diagram order \cite{EPI2016}.
The largest system size considered in this work was $L^2=256^2$ (the number of
atoms/sites is $2L^2$), with periodic boundary conditions.

%%%%%%%%%%%%%%%%%%%%%%%%%%%%%%%%%%%%%%%%%%%%%%%%%%%%%%%%%%%%%%%%%%%%%%%%%%%%%%%%%%%
{\it Semimetal-insulator transition.} To demonstrate how the BDMC technique works
and what expansion orders lead to accurate results, we benchmark it against
the semimetal - insulator (AFM) transition problem in suspended undoped ($n_e=1$) graphene.
Following Ref.~\cite{Ulybyshev2013} we introduce the ``chiral'' symmetry breaking term
\begin{equation}
H_{SB} = h \left( \sum_{{\mathbf i} \in {\mathbf A}} m_{{\mathbf i}}
- \sum_{{\mathbf j} \in {\mathbf B}} m_{{\mathbf j}} \right) \;,
\label{Stagg}
\end{equation}
where $m_{\mathbf i} = \sum_\sigma \sigma n_{{\mathbf i} \sigma}$ is the spin density operator.
[In Hybrid Monte Carlo this term is required to remove zero modes in the fermionic sector
\cite{Smith2014}; here we add it solely for the purpose of exact comparison.]
The order parameter is defined as the difference between the sublattice magnetizations:
$\ \Delta m = m_{\bf A} - m_{\bf B}$. It goes to zero when $h\to 0$
in the semimetal and saturates to a finite value in the gapped AFM phase.

\begin{figure}[tbh]
%\centering
\vspace{-1.9cm}
%\hspace{-0.28cm}
%\includegraphics[width=7cm]{GDL_Fig2.eps}
\includegraphics[width=1\columnwidth]{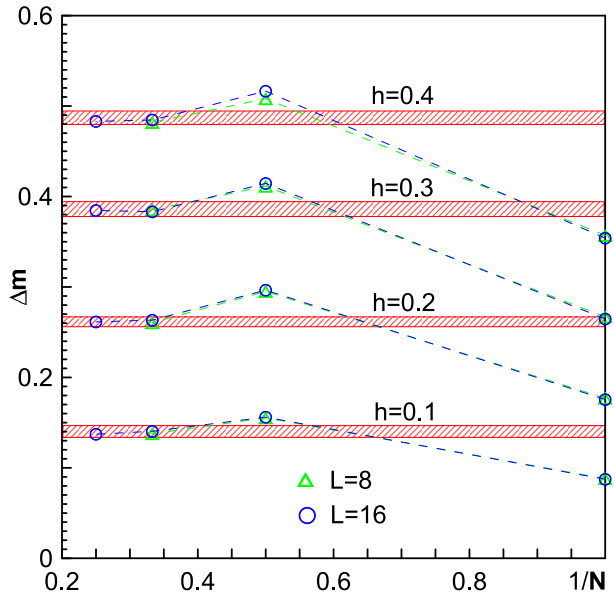}
\vspace{-2.2cm}
\caption{ (color online). The order parameter $\Delta m$ as
a function of inverse skeleton order $1/N$. Calculations were performed for system sizes $L=8$
(green triangles) and $16$ (blue circles) and for different values of the symmetry breaking field
$h =0.1, 0.2, 0.3, 0.4$. By filled red stripes we show the corresponding results (with error bounds)
from Ref.~\cite{Ulybyshev2013} for $L=18$.}
\label{Fig2}
\end{figure}

Simulations in Ref.~\cite{Ulybyshev2013} were done for
$t=2.7$, at $T=0.5$ (we use $eV$ as the unit of energy).
Screening by $\sigma$-band electrons was accounted for by adjusting on-site and n.n, n.n.n.,
and n.n.n.n coupling constants to the result of the constrained RPA calculation \cite{ScrCoul}:
they were set to $9.3$, $5.5$, $4.1$, and $3.6$, respectively. At larger distances the
long-range Coulomb potential was added as $V(r>2a/\sqrt{3})=7.2 \, a /\sqrt{3}r$ (for details
see \cite{Ulybyshev2013,ScrCoul}). The order parameter was evaluated for
$h=0.1, 0.2, \ldots $. We consider exactly the same parameter sets.

In Fig.\ref{Fig2} we compare Hybrid Monte Carlo data for $\Delta m$ to BDMC results
obtained from the skeleton expansion truncated at order $N=1,2,3,4$.
We observe that regardless of the value of $h$ the skeleton series converge, and
$N=3,4$ results are indistinguishable from results reported in Refs.~\cite{Ulybyshev2013,Smith2014}
within their error bars. We also performed calculations with the scaled,
$V(r) \to V(r)/\bar{\epsilon }$, potential (results for $\bar{\epsilon }=0.65$
are not shown here for brevity).
We confirm that gap opening takes place at $\bar{\epsilon } \sim 0.7$.
This test ensures that the BDMC
scheme is capable of capturing all important electronic correlations even in close vicinity
of the semimetal-insulator transition. While the $GW$ approximation ($N=1$) is rather
unsatisfactory in this strongly correlated regime, excellent accuracy can be reached
by extending the skeleton calculation up to order $N=3$.

{\it Dirac liquid.} We now focus on the main result of this work: the RG flow of the
effective Coulomb coupling $\alpha$ in DL described by Eqs.~(\ref{H0}) and (\ref{Hint}).
Suppressing short-range correlations by making the interaction potential flat over
some finite scale is very important
for revealing effects due to long-range forces. Otherwise, it would be impossible
to study large values of $\alpha_0$ because of the semimetal-insulator transition
taking place for unrelated reasons.

To construct the RG flow for $ \alpha (l)=\alpha_0 v^{(0)}_F/ v^{\;}_F $ we compute
Fermi velocities in systems with different linear sizes $L$ at temperatures lower
than the energy of the first excited state
$T \lesssim {\rm min} \{ \vert \epsilon_{s {\bf k}}-\mu \vert \} \propto L^{-1}$.
This is done by standard analysis of the singular part of the Green's function at the
Dirac point ${\bf K}$.
We start by obtaining eigenvectors of the $\tilde{H}=H_0+ {\rm Re} \Sigma $
operator evaluated at zero frequency ($\tilde{H}$ is a $2\times 2$ matrix in the sublattice
space for every momentum; the spin index plays no role here and is suppressed).
The corresponding eigenvalues are denoted as $\tilde{\epsilon}_{s =1,2} ({\bf k})$.
After rotating $[1 - {\rm Im} \Sigma({\bf K},\omega_n )/ \omega_n ]^{-1}_{\omega_n \to 0}$
matrix to the basis of obtained eigenvectors, we get the quasiparticle residue $Z$ from
its diagonal elements (non-diagonal elements are zero with high accuracy), and determine
the Fermi velocity from $v_F = Z [d \tilde{\epsilon}_{s} ({\bf k})/dk]_{{\bf k}\to {\bf K}}$.

\begin{figure}[tbh]
\centering
\vspace{-1.7cm}
%\hspace{-0.28cm}
%\includegraphics[width=7cm]{GDL_Fig3.eps}
\includegraphics[width=1\columnwidth]{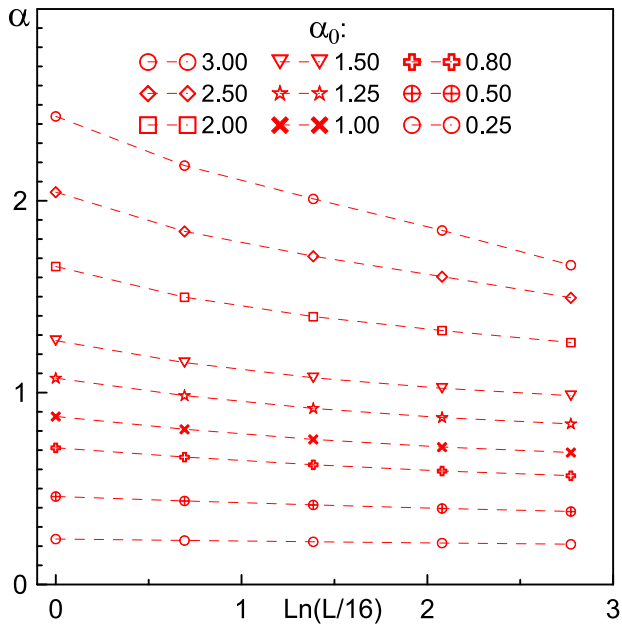}
\vspace{-1.9cm}
\caption{ (color online). RG flow of $\alpha$ with system size for various bare coupling
constants $\alpha_0$ within the GW approximation.}
\label{Fig3}
\end{figure}

In Fig.~\ref{Fig3} we present results obtained within the $N=1$, or GW, approximation for system
sizes $L=16, 32, 64, 128,$ and $256$ and a number of bare coupling constants (smaller
system sizes are disregarded to minimize the role of short-range correlations).
In all cases we find that $\alpha$ always decreases with the system size, indicating that
there is no unstable infrared point at $\alpha \sim 1$. However,  individual curves
account only for a moderate amount of renormalization and do not allow one to relate
strong to weak coupling limits by the RG flow. This deficiency is eliminated by employing
the flowgram method developed in Ref.~\cite{MasterCurve}.

The single-parameter flowgram idea is based on the assumption (to be verified by the data)
that the flow at large scales is of the form (below $\{L, \xi \} >> a$)
\begin{equation}
\alpha(L)/\alpha(\xi) = F[L/\xi] \,, \;\;\;\; F(0)=1\,,
\label{RG}
\end{equation}
where $F(x)$ is a universal function, and all dependence on microscopic parameters
is absorbed into the definition of the length scale $\xi$. If this assumption is
correct, then $[d \alpha/dl] / \alpha (\xi)$ is a universal function of $L/\xi$;
i.e., the flow derivatives (and thus the entire flow) for different microscopic
Hamiltonians have to coincide if values of $\alpha$ are matched at some large
length scale. For the logarithmic variable  $l=\ln (L/\xi )$, selecting a different
length scale $\xi$ is equivalent to shifting the flow curve horizontally.

\begin{figure}[tbh]
\centering
\vspace{-1.9cm}
%\hspace{-0.28cm}
%\includegraphics[width=7cm]{GDL_Fig3.eps}
\includegraphics[width=1\columnwidth]{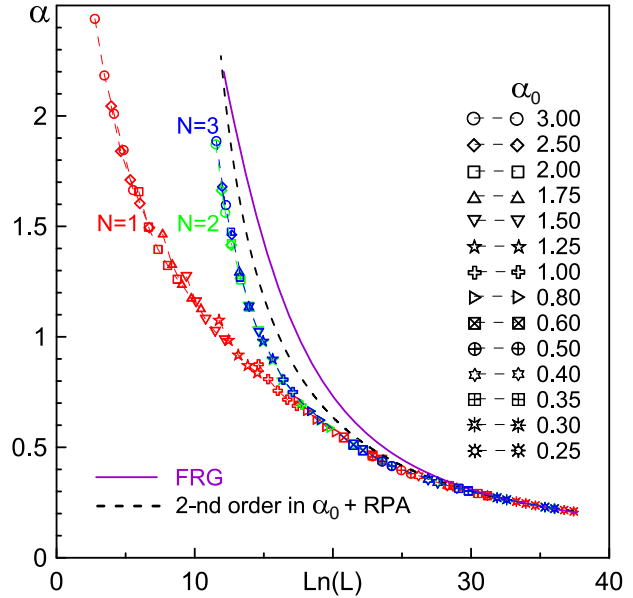}
\vspace{-2.0cm}
\caption{ (color online). RG flow of $\alpha$ with system size for $N=1$ (red), $N=2$ (green),
and $N=3$ (blue). GW curves were obtained for $L=16, 32, 64, 128, 256$; for $L=16$
there are visible deviations from the master curve because this size is not in the scaling limit
yet. Curves for $N=2$ and $3$ were obtained for $L=32, 64,$ and $128$ (see text). Black dashed
line is the RG flow based on the second-order+RPA approximation \cite{Gonzales1999,Son2007,CastrNeto2012}.
Purple solid line (courtesy of A. Sharma and P. Kopietz) is the flow obtained within the
functional renormalization group approach \cite{Kopietz2016}. }
\label{Fig4}
\end{figure}

This consideration leads to the following procedure of constructing the RG flow shown in
Fig.~\ref{Fig4}. We start with data shown in Fig.~\ref{Fig3}
and for each value of $\alpha_0$ we translate the corresponding flow curve along the $x$-axis until
$\alpha (\alpha_0,L=64)$ data overlap with the data for larger value of $\alpha_0$ at
larger $L$. Since there is no freedom in adjusting local derivatives, the scaling
hypothesis (\ref{RG}) is confirmed because all data for large enough $L$ collapse on a
single smooth master curve. The result is a flow from strong to weak coupling that
effectively extends over twelve (!) orders of magnitude.

After establishing the RG flow in the $GW$ approximation, we repeat the above
analysis for higher-order skeleton formulations with $N=2$ and $N=3$.
Since these computations are more demanding they were mostly limited to
$L=32$ and $64$ (and $L=128$ for smaller values of $\alpha_0$ at $N=2$).
The protocol of constructing the master curves for $N=2, \; 3$ is exactly
the same as for $N=1$, i.e. it is obtained by shifting flow curves horizontally. The result
is shown in Fig.~\ref{Fig4}. Clearly, the effect of vertex corrections is very pronounced
for $\alpha >1$, but the skeleton sequence quickly converges and $N=3$ results are nearly identical
to those for $N=2$. The final flow is always to the asymptotically free DL with logarithmically
divergent Fermi velocity and finite quasiparticle residue.

%%%%%%%%%%%%%%%%%%%%%%%%%%%%%%%%%%%%%%%%%%%%%%%%%%%%%%%%%%%%%%%%%%%%%%%%%%%%%%%%%%%
{\it Conclusions.} We address the fundamental problem of the Dirac liquid stability against
strong Coulomb interactions using the Diagrammatic Monte Carlo method that allows us
to account for higher-order vertex corrections within the fully self-consistent
skeleton expansion. We find that even in the strongly correlated Dirac liquid
the skeleton sequence quickly converges and leads to an accurate solution of the
RG flow for the effective coupling constant. The unstable infrared point at
$\alpha \approx 1$ is ruled out, and the flow is found to be always to the asymptotically
free state.

Our approach is general and can be applied to any graphene-type system with arbitrary
dispersion relation featuring Dirac cones, both doped and undoped, and with arbitrary shape of
the interaction potential. Given that long-range electron-ion interactions in these
systems are of the same Coulomb origin, future work should address them as well to achieve the
best effective description of realistic materials.

%%%%%%%%%%%%%%%%%%%%%%%%%%%%%%%%%%%%%%%%%%%%%%%%%%%%%%%%%%%%%%%%%%%%%%%%%%%%%%%%%%%
{\it Acknowledgements.} We thank B. Svistunov for discussions. This work was supported by
the Simons Collaboration on the Many Electron Problem, the National Science Foundation under
the grant PHY-1314735, the MURI Program ``New Quantum Phases of Matter" from AFOSR, the
Stiftelsen Olle Engkvist Byggm\"{a}stare Foundation, and the Swedish Research Council grant
642-2013-7837.
%%%%%%%%%%%%%%%%%%%%%%%%%%%%%%%%%%%%%% BIBLIOGRAPHY %%%%%%%%%%%%%%%%%%%%%%%%%%%%%%%%%%%%

\end{document}